# SiN-on-SOI Optical Phased Array LiDAR for Ultra-Wide Field of View and 4D Sensing

Baisong Chen[1], Yingzhi Li[1], Qijie Xie[2], Quanxin Na[2], Min Tao[1], Ziming Wang[1], Zihao Zhi[1], Heming Hu[1], Xuetong Li[1], Huan Qu[1], Yafang He[1], Xiaolong Hu[1], Guoqiang Lo[3] and Junfeng Song[1,2,*]

[1]*State Key Laboratory on Integrated Optoelectronics, College of Electronic Science and Engineering, Jilin University, Changchun, 130012, China*
[2]*Peng Cheng Laboratory, Shenzhen, 518000, China*
[3]*Advance Micro Foundry Pte. Ltd., 11 Science Park Road, Singapore Science Park II, Singapore, 117685*
*Corresponding author: songjf@jlu.edu.cn

**Three-dimensional (3D) imaging techniques are facilitating the autonomous vehicles to build intelligent system. Optical phased arrays (OPAs) featured by all solid-state configurations are becoming a promising solution for 3D imaging. However, majority of state-of-art OPAs commonly suffer from severe power degradation at the edge of field of view (FoV), resulting in limited effective FoV and deteriorating 3D imaging quality. Here, we synergize chained grating antenna and vernier concept to design a novel OPA for realizing a record wide 160°-FoV 3D imaging. By virtue of the chained antenna, the OPA exhibits less than 3-dB beam power variation within the 160° FoV. In addition, two OPAs with different pitch are integrated monolithically to form a quasi-coaxial Vernier OPA transceiver. With the aid of flat beam power profile provided by the chained antennas, the OPA exhibits uniform beam quality at an arbitrary steering angle. The superior beam steering performance enables the OPA to accomplish 160° wide-FoV 3D imaging based on the frequency-modulated continuous-wave (FMCW) LiDAR scheme. The ranging accuracy is 5.5-mm. Moreover, the OPA is also applied to velocity measurement for 4D sensing. To our best knowledge, it is the first experimental implementation of a Vernier OPA LiDAR on 3D imaging to achieve a remarkable FoV.**

## 1 Introduction

Driven by artificial intelligence, autonomous systems such as autonomous driving and smart warehouse are making a profound impact on society. As a remote sensing technology, light detection and ranging (LiDAR) with dense point cloud images of the real-world is a crucial technique for autonomous systems. Traditional LiDAR system based on mechanical configuration faces challenge from solid-state LiDAR characterized by strong robustness, high performance and potentially low cost. The solid-state schemes mainly include focal plane array (FPA), focal plane switch array (FPSA) and optical phased array (OPA). There are two main schemes for ranging by using FPAs reported. One is 905-nm based time-of-flight (TOF) scheme combined with single-photon avalanche diodes detector (SPAD) array[1-3]. The other one is coherent detector array focusing on 1550 nm[4-6]. The TOF FPAs realize the compact integration of SPAD and readout circuit through the 3D-stacked CMOS technology[3], showing potential application in miniaturized mobile device. Nevertheless, the excessive sensitivity to ambient light posed the SPAD a risk when it operates in strong light scene. The coherent FPAs[4-6], as reliable 3D sensors, are insensitive to ambient light and can simultaneously detect the distance and velocity of the objects. Rogers et al. have demonstrated a 3D imaging system of coherent FPA in 2021, achieving 3D imaging and velocity measurement[6]. However, limited FoV and the pixel numbers may not meet the requirement for the autonomous driving. The FPSAs can be mainly divided into solid-state Mach-Zehnder interferometer (MZI) switches and semi-solid-state microelectromechanical systems (MEMS). The MZI-FPSAs chip can only accommodate small number (e.g., tens) of pixels due to the footprint of the thermal MZI[7,8]. Boosted by the low power consumption and small footprint of MEMS switch, Zhang et al. has demonstrated an ultra-large FPSAs with 16,384 pixels[9]. However, MEMS FPSA scheme is not fully solid-state, so there are still concerns of reliability. Besides, wide FoV 3D imaging has not been demonstrated.

OPA, as a promising candidate for fully integrated solid-state LiDAR, can perform wide-angle and agile horizontal beam steering[10-19]. Moreover, compared with the aforementioned focal plane scheme, OPAs can directly manipulate the wavefront to achieve on-chip beam shaping without any lenses or optomechanical structures. The OPA can be beneficial to 3D imaging, especially in the application of autonomous driving requiring greater than 120°. To achieve wide imaging FoV, two considerable factors are the aliasing-free FoV[14,19] and the far-field profile[16,20]. Beam aliasing-free FoV ensures a sole main beam to exist in a FoV without interference by undesired beams. A wide far field profile can provide consistent main lobe power over a wide FoV for high-quality 3D imaging. The far-field profile can be optimized through the specific design on the grating antenna.

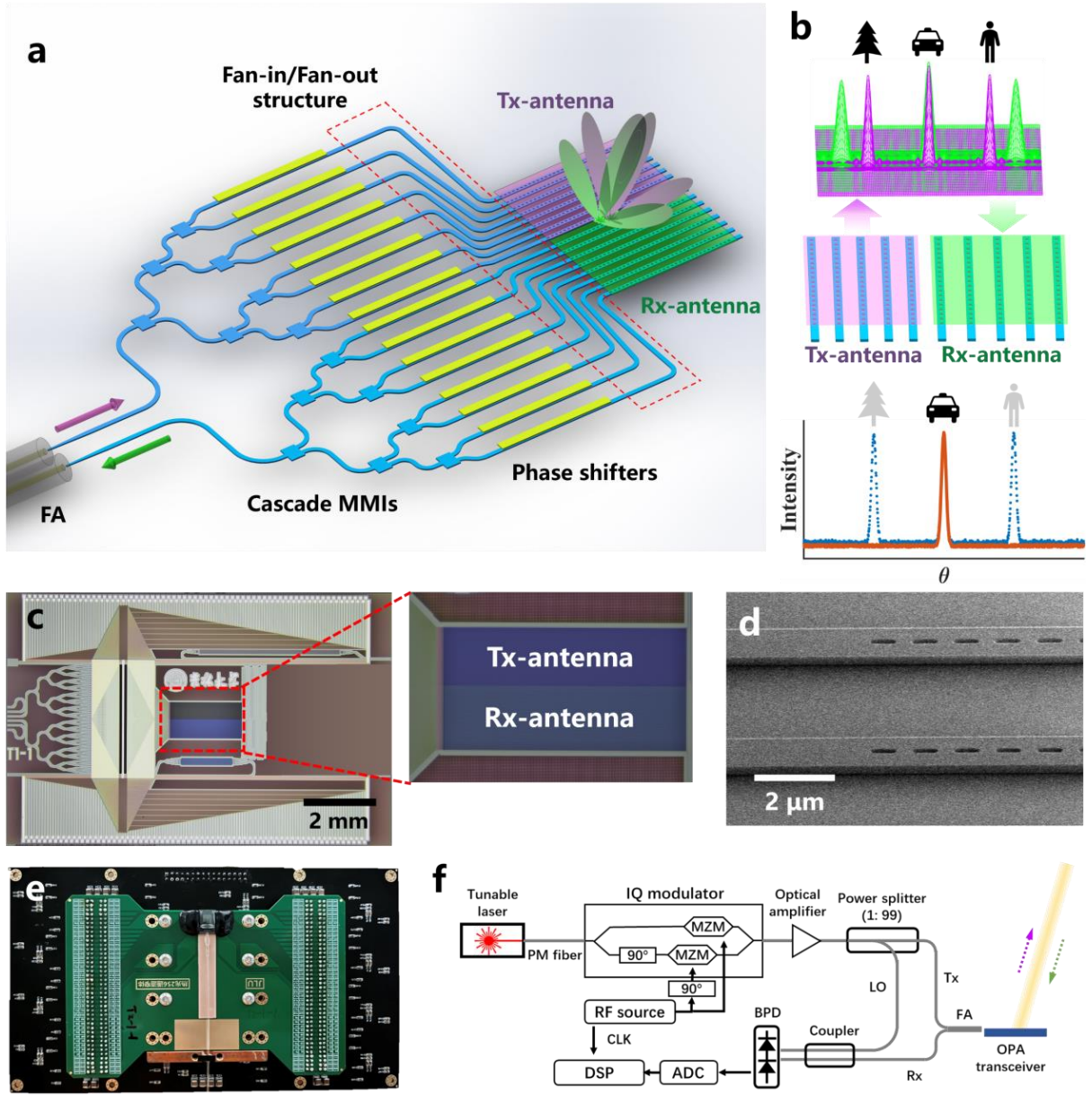

**Fig. 1 The silicon-based Vernier OPA structure and OPA-based FMCW LiDAR a** Schematic for the structure of Vernier OPA with chained grating antennas. The transmission and reception of light are realized by two monolithically integrated OPAs. The tightly arranged antennas form a quasi-coaxial system, isolating undesired interference in the optical link. **b** The ranging concept of a Vernier-OPA-based LiDAR. The main lobes of Tx- and Rx-OPAs are aligned in one direction, and other grating lobes are misaligned. Consequently, the aliasing signal can be avoided completely and thus the distance of target can be resolved accurately. **c** Optical microscope image of the fabricated OPA chip. The Tx- and Rx-antenna are shown in detail in the inset **d** Scanning electron microscope (SEM) image of the large profile grating antennas. **e** The packaged OPA chip with aligned fiber array and a multi-channel DAC driving circuit. **f** Schematic for the OPAs-embedded FMCW LiDAR system.

OPAs with 180° aliasing-free FoV have been extensively studied[15,16,19,21]. Index-mismatch waveguide[21] and slab grating [15,19] are typically used to reduce the crosstalk between adjacent waveguides, achieving 180° aliasing-free FoV. However, the narrow far-field profile determined by grating antenna shrinks the available range of FoV. Alternatively, non-uniform antennas arrays can also achieve wide beam aliasing-free FoV[16,17]. However, the efficiency of the main beam needs to be further improved. Poulton et al. demonstrated a record-performance 8192-element OPA with flip-chip attached ASICs to achieve the first dense 3D point cloud based on OPA[11]. By the merit of 1-μm antenna pitch, the scheme realizes 3D imaging with a 100° FoV, which demonstrated the largest reported FoV for OPA-based LiDAR so far. The performances of reported OPAs are summarized in supplementary information Table S1.

In this work, we present a SiN-on-SOI OPA based LiDAR, which demonstrates a record wide imaging FoV of 160° for 3D imaging. First of all, we utilize the chained grating to construct OPA antenna to broaden the far-field profile. Consequently, the OPA exhibits less than 3-dB beam power variation within 160° FoV, which provides sufficient beam power even at the edge of the FoV and thus guarantee the 3D imaging quality uniformly. Furthermore, we integrate Tx- and Rx-OPA with different grating antenna pitch to form a 256-element Vernier OPA, aiming at extension of aliasing-free FoV. The sufficient grating pitch offers great flexibility in the grating antenna design to improve the OPA performance significantly. In particular, Tx- and Rx-antenna arrays are placed sufficiently close to form a quasi-coaxial system, relaxing the burdens of optical alignment between the transmitter and receiver[6]. Moreover, the separation of the transmitted and the received signals on the optical path can help avoiding undesired interference from interface reflection[22]. Lastly, we apply the OPA in the FMCW LiDAR system and demonstrate a wide scene, high-resolution 3D imaging with approximately 7000 points and a record 3D imaging coverage exceeding 160°. To the best of our knowledge, the demonstrated FoV has been the widest scene for OPA-based LiDAR in the 3D imaging application so far. Furthermore, the FMCW LiDAR system is applied to 4D sensing for velocity measurement and high-precision ranging. N. DOSTART has reported the proof-of concept Vernier OPA transceiver simply based on two antenna arrays[23], but large angle steering and 3D imaging have not yet been achieved. This work is the first demonstration of a well-designed Vernier OPA based LiDAR enabling arbitrary beam scanning. Moreover, the chained grating antenna provides a flat and broad beam power profile. Simultaneously, by virtue of the effective suppression on grating lobes via vernier structure, the OPA can eliminate beam aliasing over ultra-wide FoV. Therefore, the proposed OPA LiDAR exhibits a remarkable FoV in 3D imaging with almost no blind spots. It facilitates significantly the application of integrated OPA chip in autonomous driving, as the safety and wide scene are highly demanded.

## 2 Principle verification and structural design

Fig. 1a shows the schematic of the proposed Vernier OPA transceiver. A 128-channel Tx-OPA and a 128-channel Rx-OPA are integrated monolithically into the transceiver. Tx-OPA and Rx-OPA with similar structures are used for light transmission and reception, respectively. The OPA mainly include edge coupler, cascaded multimode interference (MMI) couplers, SiN-Si transition, thermal-optical (TO) phase shifters and grating antenna arrays. An incident light is coupled into the chip through the edge coupler. The cascaded MMIs divide the light evenly into 128 waveguides, and then the phase of each waveguide is controlled individually by the corresponding thermal-optical phase shifter. At the output of the phase shifter arrays, the position of waveguide array is adjusted to match the 3.8 μm Tx-antenna array emits the light into the free space. By controlling the phase of each channel and input light wavelength, the OPA can agilely achieve horizontal and longitudinal beam steering. The reflected light is received by the Rx-antenna array with 4.2-μm waveguide pitch. The light is then directed out through a fan-out structure and coupled to the off-chip photodetector. The proposed monolithic Vernier OPA with 256 channels (128 channels for Tx-OPA and 128 channels for Rx-OPA) is fabricated on a multilayered SiN-on-SOI (silicon on isolator) platform. We use 1 μm-wide silicon nitride (SiN) waveguide with thickness of 0.34 μm. Large-aperture SiN dual-level grating antenna arrays can perform unidirectional emission[24], achieving ~90% unidirectionality. The divergence angle of OPA at 0° steering angle is 0.16°×0.04° (Simulation and measurement details in supplementary Fig. S1 and S2). The tuning efficiencies of phase shifters are measured to be around 6 mW/π with the suspended structure[25].

The operation principle of the Vernier antennas is shown in Fig. 1b. For uniform antenna array, when the pitch of [23]grating antennas is greater than half a wavelength, the grating lobes rise in the FoV inevitably[26]. However, the direction of grating lobes can be manipulated via controlling the pitch among the grating antennas. Therefore, when the target beams of Tx- and Rx-antenna are aligned in one direction, other aliasing grating lobes will be misaligned. In this case, free aliasing can be achieved within the FoV by using the Vernier OPA structure. The Vernier grating antenna can effectively distinguish the target direction by the generating a beat signal with high signal-to-noise ratio, leading to enhance ranging performance, as shown in Fig. 1b. The optical microscope image of the fabricated OPA is shown in Fig. 1c. The zoom-in image of the Vernier grating antennas is exhibited in the inset. The tightly arranged Tx- and Rx-antennas enable OPA-based LiDAR to operate in quasi-coaxial system, relaxing the complicated optical alignment. An aliasing-free FoV is achieved through the vernier antenna arrays without any limitation in grating structure, that allows various grating structure.

In order to further expanding the imaging FoV, we design a chained grating antenna that can emit a broad far-field profile with efficient and stable beam power over the FoV. It is beneficial to improve the imaging quality at edge of FoV. The Fig. 1d is the SEM image of the designed chained grating antenna. As shown in the Fig. 1e, the OPA chip is packaged with polarization-maintained fiber array (PMFA). The chip is wire-bonded to circuit board, and controlled by a 256-channel DACs driver[27]. Each phase shifter is driven by a digital-to-analog converter (DAC) to implement phase control. Thereby, arbitrary wavefront modulation and beam scanning can be realized by electrical control. The solid-state 3D-imaging system realized by applying the Vernier OPA to the FMCW LiDAR system, is shown in Fig. 1f. A chirped laser is generated by externally modulating a tunable laser through an in-phase/quadrature modulator (IQM). The linear frequency chirp is synthesized by a radio frequency (RF) module. The light beam from the amplified chirp laser is emitted and steered by the OPA. Finally, the beat signal of received light and local light is detected by a balance photodetector (BPD) followed by the processing by an analog-to-digital converter (ADC), and finally solved by digital signal processing (DSP).

In the case of OPA design, the peak power of the steering beam within FoV is modulated by the far-field profile of a single grating antenna. Hence, the peak power of beam normally decrease as the steering angle increased[16,19]. To obtain a broad far-field profile with a smooth beam power distribution, we propose chained grating antennas. By simply reducing the etching aperture, the size of the emitter can be reduced and thus the far-field profile can be expanded effectively. Compared with using thick Si waveguide (0.4 μm) to improve optical confinement and reduce the size of emitters[20], chained gratings have a simpler process and can more effectively broaden the far-field profile. Here, we apply the chained grating antennas to the OPA design. The simulated far-field profile in the case of 200-nm etching width is shown by the blue line in Fig. 2a. The 3-dB width is about 115°. The calculated far-field profiles of different etching width can be referred to the supplementary Fig. S3.

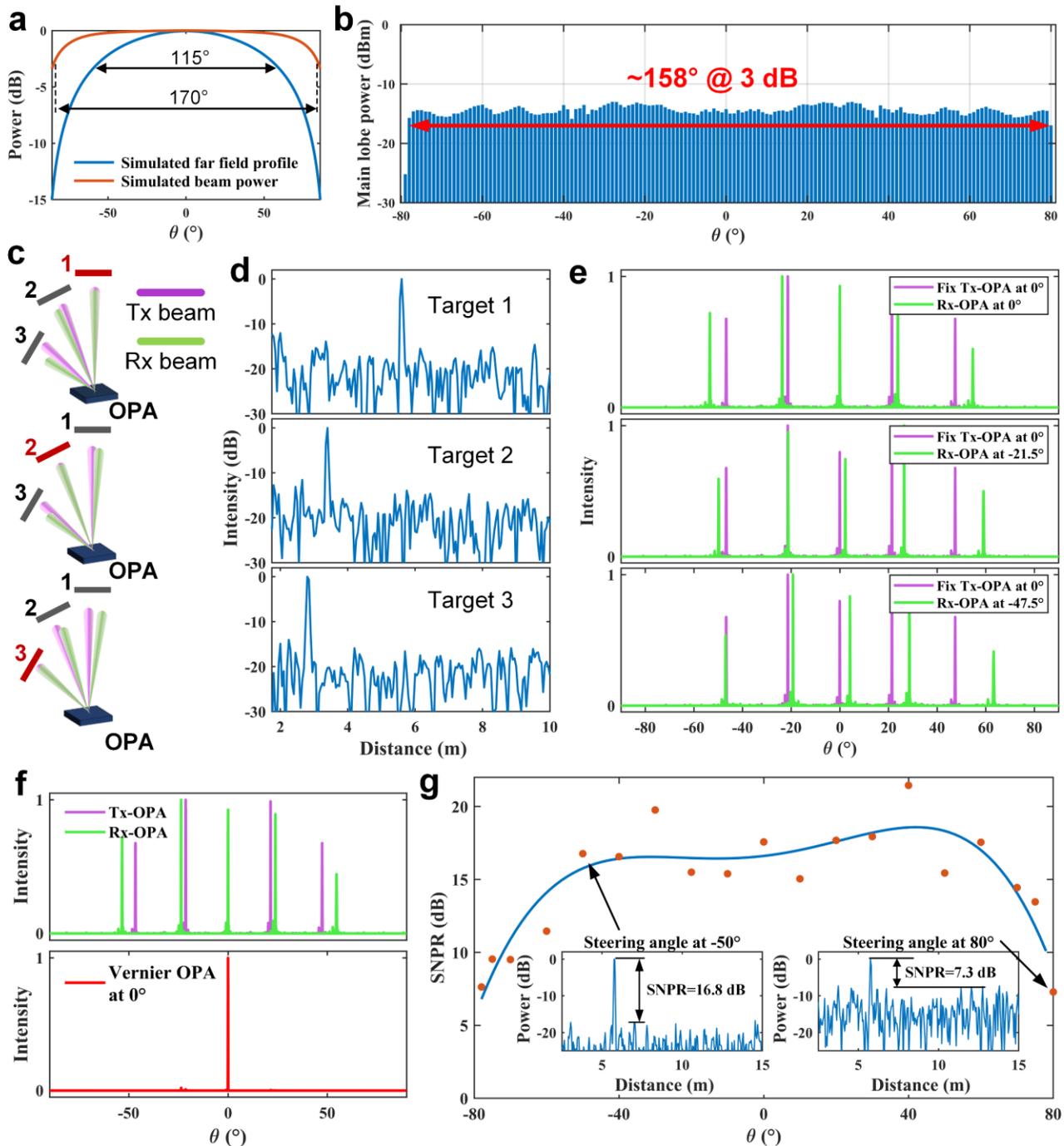

**Fig. 2 Characterization of the Vernier OPA with a wide imaging FoV. a** Simulated far field profile and beam power. **b** Measured beam power of the fabricated OPA within -79° to 80° with input power of 0 dBm, and the beam power variation is less than 3 dB within the FoV of 158°. **c** Setup for verifying the vernier OPA. Three targets are arranged at 0°, -21.5° and -47.5° for the lobes of Tx-OPA. **d** Beat signals of three targets by steering Rx-OPA to the target beam of Tx-OPA. **e** Measured far-field of Tx- and Rx-OPA over 180°. The main lobe of Tx-OPA is fix at 0°, and the grating lobes point to ±21.5° and ±47.5°. The Rx-OPA is aligned to the main lobe of 0° and grating lobes of -21.5° and -47.5° of Tx-OPA, sequentially. **f** Measured results for far fields of Tx-OPA, Rx-OPA (upper) and the corresponding Vernier OPA (lower) when the steering angle is 0°. **g** The plot for signal-to-noise pedestal ratios (SNPR) of the beat signals against as the steering angle within 160° FoV. The orange dots represent the experimental measurement and the blue line represents the fitting curve. The insets are the beat signals for ranging as the OPA steered at -50° and 80°, respectively.

The calculated beam power along with the steering angle is shown by red line in Fig. 2a. The 3-dB width is close to 170°. Since the total beam power is the power integration over the beam width[26], the power degradation at the edge of far-field profile can be balanced by the broadened divergence angle of the light beam. Therefore, a broad far-field profile with a slow roll-off power gives rise to a wide effective FoV for 3D imaging. The experimental result for the total beam power steered from -79° to 80° with an interval of 1° is shown in Fig. 2b. The measured profile of total beam power matches well with simulation result. The power drop at the -79° and 80° is caused by the obstacle stuck on both sides of the OPA chip. It is found that the fabricated OPA exhibits less than 3 dB beam power variation within a record ~160° FoV. In fact, a horizontal FoV close to 160° can already meet the vast majority of applications. More details of the beam quality within the FoV are demonstrated in the supplementary information Fig. S4.

Next, the characterization of the Vernier OPA is conducted. According to our design, the target beam of the fabricated Tx-OPA is fix at 0°, while the ±1st and ±2nd order grating lobes are located at approximately ±21.5° and ±47.5°, respectively. In order to verify the aliasing-free detection operation of the vernier transceiver, three targets are placed at 0°, -21.5° and -47.5° at different distance, respectively as shown in Fig. 2c. The Rx-OPA is electrically controlled to aligned to the fixed Tx-OPA in the three directions sequentially. The distances of three target are resolved based on the beat signals shown in Fig.2d. It is found that the noise from the grating lobes suppressed by the Vernier OPA significantly and only a clear beat signal appears during the ranging process. The measured normalized far fields of fabricated Tx- OPA and Rx-OPA over 180° are obtained by splicing images captured by infrared camera, as shown in Fig. 2e. When the target beam of Rx-OPA is sequentially aligned to the main lobe, -1st and -2nd grating lobes of the Tx-OPA, the rest of beams are staggered except for the alignment direction. In addition, the superimposition of measured far fields for Tx-and Rx-OPA is shown in the upper of the Fig. 2f. Both of the Tx-and Rx-OPA are steered at 0°. The corresponding far field for Vernier OPA is shown in the lower of the Fig. 2f. The resultant far field of Vernier OPA is synthesized by overlapping the far fields of Tx- and Rx-OPA. Noted that only the target beam spikes within the FoV and the rest of grating lobes are suppressed strongly. The simulated and measured results for far fields of fabricated Vernier OPA at different steering angles are shown in the supplementary Fig. S5 and Fig. S6.

To clearly obtain the distance information for 3D imaging over an ultra-wide FoV, the beat signal at ranging measurement requires sufficiently high signal-to-noise pedestal ratios (SNPR). The broad far-field profile provided by the chained grating antennas can facilitate the Vernier OPA-based LiDAR to generate high-SNPR beat signal. We measure the SNPR of beat signals within a ~160° FoV to evaluate the ranging performance of the OPA, as shown by the orange dots in Fig. 2g. It is found that the SNPR of beat signals can be maintained above 10 dB when the steering angle is from -60° to 70°. The blue curve shown in Fig. 2g is a fitting for the measured SNPR. The fitting curve of SNPR exhibits flat top from -60° to 70° steering angle, attributing to the flat beam power profile of chained grating antennas. The uniform beam quality lies profound foundation for high quality 3D imaging with ultra-wide FoV. Although the drop of SNPR occurs at the both edges of the FoV, the beat signals still overtop the noise pedestals by 7.5 dB sufficiently high for 3D imaging. The degradation in the SNPRs of beat signals at the edges of the FoV is probably caused by the shrunk effective aperture of Rx-antenna at a grazing angle. In addition, the RF spectra for beat signals at the -50° and 80° steering angle are shown in the insets of Fig. 2g., respectively. To sum up, the effective FoV provided by the proposed OPA-based LiDAR for 3D imaging can reach to 160°. The measurement setup and beat signals at different steering angles are demonstrated in the supplementary information Fig. S7.

## 3 3D imaging and velocimetry

Firstly, the solid-state OPA LiDAR system is applied to measure velocity and distance as shown in Fig. 3. The bandwidth for the linear optical frequency chirp is 3 GHz, and the period for the up- and down-chirp time is 100 μs. The coherent LiDAR can directly measure the target velocity through Doppler frequency shifts of the received light[28]. The velocity-annotated 4D point clouds image of a rotating wheel is illustrated in the Fig. 3a. The Fig. 3b shows the measured beat signals for a static wheel and rotating wheel. When the wheel stops, the beat signal of the up-chirp is consistent to that of down-chirp, indicating the measured distance is 3.9 m. When the wheel starts rotating, due to the Doppler shift of reflected light, the up- chirp and down-chirp signal shift in the opposite direction, and the median of the two signals is the ranging distance of 3.9 m, and the difference of them can be solved as velocity of 0.217 m/s. We utilize the OPA to steer the light beam along the edge of the rotating wheel. The distance of 16 positions and corresponding projection of velocity in the direction of beam propagation are measured, as shown in Fig. 3c. The velocity projection along the edge of the rotating wheel varies linearly from -0.3 m/s to 0.3 m/s. It is also found that there is a jitter in the outline of the wheel in the point clouds. It may be attributed to the instability of the system caused by the nonlinearity of chirp modulation and the linewidth of the laser[29,30]. To quantify the accuracy of ranging, we conduct a 3D imaging of cardboard boxes with continuous surfaces, and measure three positions on cardboard boxes by 100 times as shown in the supplementary information Fig. S8. The standard deviation of the distance measurement is approximately 5.5 mm.

We also apply the OPA-based FMCW LiDAR system to distance measurement. A long optical path is constructed by placing a mirror in front of the front wall, as shown in Fig. 3d[10,16]. The OPA-based LiDAR can measure the distance to the back wall through the mirror. The 3D point clouds of the front wall, mirror mount and image of back wall are shown in Fig. 3e. The point clouds of the front and back walls are demonstrated in Fig. 3f and g respectively to display details. The mirror mount is clearly observed in the point clouds and significantly distinguished from the front wall. The contour of the back wall point clouds is that of the mirror. The beat signals of ranging front wall and back wall are shown in Fig. 3f and g, and the distances of front wall and back wall can be resolved as 6.75 m and 22.15 m, respectively. In the experiment, the input power to the OPA is 25 dBm and thus a ~11 dBm main beam is emitted from the OPA. The SNPRs of the generated beat signals for targets at 6.75 m and 22.15 m are approximately 14 dB and 9.5 dB, respectively. Furthermore, photonics integrated circuit based on SiN-on-SOI platform can withstand input power exceeding 37 dBm[11], enabling longer ranging.

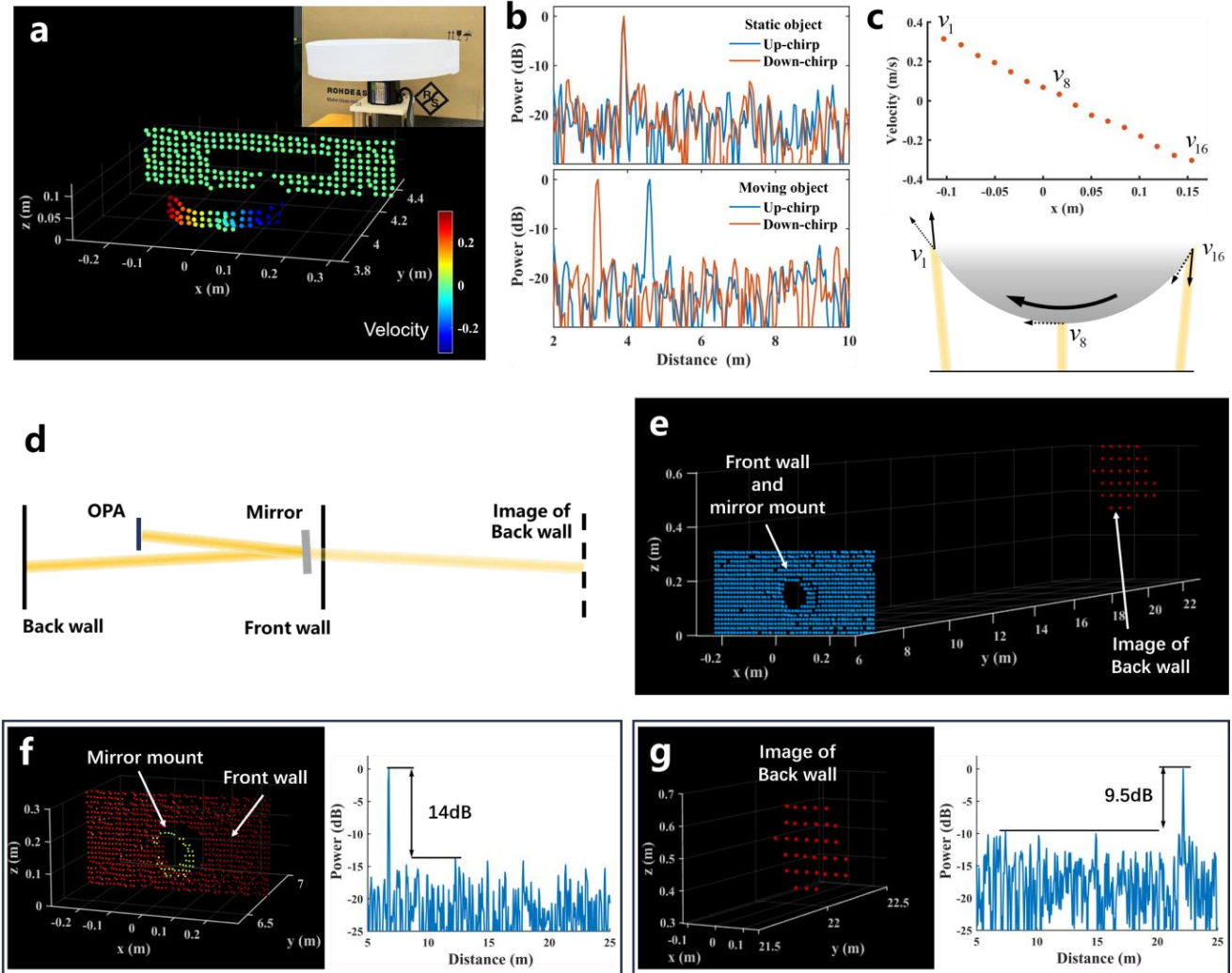

**Fig. 3 3D imaging results. a** Velocity-annotated point clouds and camera image of a rotating wheel. **b** The measured beat signals of static and moving object, respectively. **c** Axial velocity across the middle of the rotating wheel, and the below illustration shows the projection of the velocity of the wheel onto the scanning beam. **d** Setup for long range through folding the optical path by a mirror. **e** 3D point clouds of the front wall and mirror mount at about 6.75 m and image of back wall at 22.15 m. **f** and **g** Point clouds of the front wall, mirror mount and back wall as well as the representative beat signals, respectively.

Last but not least, we demonstrate the OPA-based solid-state LiDAR with an ultra-wide FoV in 3D imaging. A high-resolution 3D point clouds exceeding 135° with more than 7000 resolvable points including 400 horizontal resolvable points and 18 vertical resolvable points as shown in Fig. 4a. The inset shows the corresponding photo of the real scene. Noted that the horizontal angle of the proposed OPA-LiDAR is not limited to 135°. Furthermore, we have demonstrated the horizontal steering angle of the OPA-based LiDAR can be improved to achieve 3D imaging with a record wide 160° FoV. The corresponding 3D point-clouds image is illustrated in the Section 9 of the Supplementary information. Limited by the bandwidth of Erbium-Doped Fiber Amplifier (EDFA), the vertical FoV is confined to ~3° within a wavelength range of 42 nm. A wider bandwidth optical amplifier can enable a broader vertical FoV to be achieved[10].

The 3D point cloud clearly exhibits a series of letters 'JILIN UNIVERSITY' located at the middle area of the FoV, with a white curtain wall on the right and scattered cardboard boxes on the left. OPA owns flexible angular resolution, which can provide more detailed 3D images by increasing the angular resolution. Complex objects require extremely high angular resolution scanning. For objects with simple outlines, such as walls, moderate resolution can also clearly display the outlines. Therefore, the area where the letters are located is scanned with a 0.25° angle resolution to better display details, while the walls and cardboard boxes with simple outlines are scanned with a 0.5° angle resolution. The outline of the letters can be clearly distinguished, and the shadows of the letters on the back wall are also clearly presented. The point clouds image of the letter 'R' is enlarged and shown in Fig. 4b. The outline of the letter R is clearly delineated by approximately 100 points. The ripples of the window curtains are clearly displayed in the 3D point clouds, as shown in Fig. 4c. As shown in Fig. 4d, the point clouds of the right curtain wall are also specially displayed. The gradual change in distance from about 2.5 to 4.2 m can be accurately observed through the changes in colors. By virtue of the uniform beam power profile supported by the proposed chained antenna, the OPA based LiDAR exhibits superior and uniform quality of 3D imaging over the whole FoV. As shown in Fig. 4 (a), we can barely observe blind spot in the point clouds over 135° horizontal FoV. To our best knowledge, this is the first time that OPA solid-state LiDAR has demonstrated 3D imaging exceeding 160°. The OPA-based FMCW LiDAR system can potentially process more than the demonstrated 7000 points in 3D imaging.

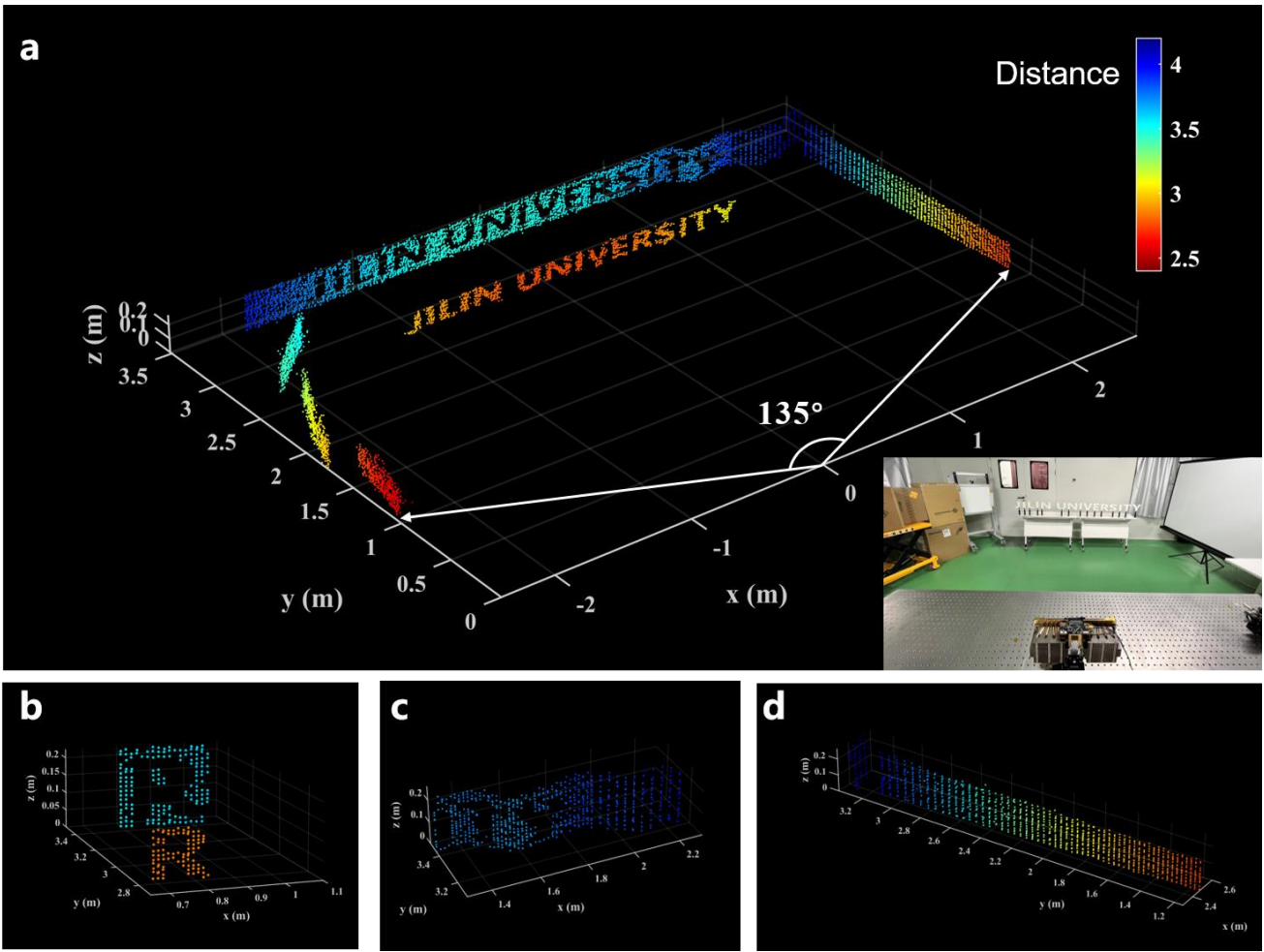

**Fig. 4 3D point clouds construction for a wide-scene scenario. a** Point clouds with a FoV of over 135°, and the inset is the camera image. **b** Point clouds in detail of the letter 'R'. **c** Point clouds in details of window curtain. **d** Point clouds of the right wall showing the gradual change in distance.

**Table 1 Performance comparison of on-chip (semi-) solid state platform for 3D imaging**

| Year | Mechanism | Tx/Rx alignment | Integration | Horizontal imaging FoV |
|---|---|---|---|---|
| 2019[10] | OPA | Non-coaxial | Separated chips | 56° |
| 2020[6] | FPA | Non-coaxial | Separated chips | - [a] |
| 2022[9] | MEMS-based FPSA | Coaxial | Monolithic | 16° |
| 2022[11] | OPA | Coaxial | Monolithic | 100° |
| **This work** | **Venier OPA** | **Coaxial** | **Monolithic** | **160°** |

[a] No accurate data available.

## 4 Discussion

The comparison among various on-chip solid state (or semi-solid state) platform for 3D imaging is illustrated in Table 1. It is shown that the proposed Vernier OPA-based LiDAR system exhibits outperformance in term of monolithic integration, easy optical alignment and wide imaging FoV. In this work, we have demonstrated a solid-state OPA-based LiDAR with a record wide horizontal FoV of 160° for 3D imaging. The chained grating antenna is proposed to broaden the far-field profile of the OPA. Therefore, the beam power variation of OPA is less than 3-dB within a 160° FoV, providing sufficient beam power even at the edge of FoV. On the other hand, two chained grating antenna arrays with different grating pitch are integrated monolithically to form a 256-channel Vernier transceiver, leading to an aliasing-free full FoV. The architecture of combining wide far-field profile antenna and Vernier OPA enables the OPA based LiDAR to prove outstanding 3D imaging quality with an impressive wide horizontal FoV of 160°. Alternatively, the amounts of resolvable points at horizontal and vertical dimensions can be scaled up by finer phase control and wavelength tuning. It can guarantee high-angle-resolution beam steering. To the best of our knowledge, it has been the widest FoV for 3D imaging based on OPA-based LiDAR so far. In addition, the Vernier OPA is also applied to the FMCW ranging system for velocity and distance measurement. The velocity-annotated 4D point clouds and distance measurement up to 22 m indoors are demonstrated. The precision of ranging is 5.5 mm.

It is worth of mentioning that the closely spaced Tx- and Rx-antenna is beneficial to the optical imaging system. In particular, the Tx- and Rx-antennas are located close to form a quasi-coaxial system, avoiding complex alignment between transmitter and receiver[6,10]. Moreover, the separation of transmitted and received light in the link helps to avoid unexpected interference from interface reflection. The entire system does not contain any lenses or optical mechanical structures, thus can be miniaturized to a compact, low cost and high-reliability platform.

The SOI platform has commonly been a competitive alternative for OPA fabrication, since strong optical confinement for dense grating antenna arrays. Hence, the aliasing free FoV can be expanded effectively. However, the power handling capability of silicon is less sufficient for the application in OPA system. Rather than relying on material characteristics, we propose a novel architecture for OPA chip design to achieve wide imaging FoV with a flat beam power profile. The proposed structure is suitable to various material platforms for the fabrication of high performance OPA[31,32].

In this work, a fabricated OPA chip with 256 elements is used to demonstrate 3D imaging of indoor scene at only around 22 m. Nevertheless, by virtue of advanced SiN-on-SOI platform and high-power handling of SiN waveguides, LiDAR for longer distance can be achieved by enlarging the OPA scale of channels and increasing input power. The range of LiDAR scales as the square root of the antenna aperture and transmission power[33]. Therefore, our architecture can potentially operate at ranges of exceeding 200 m by scaling the proposed monolithic OPA transceiver to thousands of elements and boosting the input power to more than 30 dBm.

## 5 Appendix
**Simulation**

Simulation results of the far-field profile of grating antenna are done by FDTD (ANSYS LUMERICAL, USA). The simulation of the far fields of OPAs are calculated by the superposition of the diffraction fields generated by antennas.

**Measurement**
1. Beam power within the FOV

The main beam power is measured by a free space power meter (S122C, Thorlabs, USA). The OPA and controller are placed on the rotation stage and a power meter is placed in a fixed position. By controlling the rotation of rotation stage and the beam steering of OPA, beams of different angle are aligned to the power meter to obtain the power of each steering angle in the FOV.

2. Far fields of Vernier OPA

The far fields of Tx- and Rx-OPA are measured by an infrared camera (Bobcat 640 GigE, Xenics, Belgium). Due to the limited FOV of the infrared camera, the far fields of OPA in full FOV cannot be directly obtained. We place the OPA on a motorized rotation stage and place the observation screen properly. The infrared camera is adjusted to exactly cover the 10° range of far fields of OPA. We rotate the rotation stage with 10° intervals, capture images through the camera, and finally splicing the images to obtain the far fields of Tx- and Rx-OPA in the full FOV. The resultant far fields of Vernier OPA are synthesized by overlapping the measured far fields of Tx- and Rx-OPA.

3. OPA-based LiDAR system

A tunable laser (TSL550, Santec, Japan), an IQ modulator and a chirp-frequency-modulation RF module are used to form a FMCW laser. The chirp bandwidth is 3GHz, and the up- and down-chirp lengths are 100 μs. The Vernier OPA transceiver completes emission of light and reception of reflected light. The beat signals of received light and local light are sampled by an oscilloscope (MSO64B, Tektronix, USA), and finally solved by DSP in PC software.

## Acknowledgements
This work is supported by The National Natural Science Foundation of China under Grants No. 62090054, No. 61934003, No. 62105174 and No. 62105173; National Key R&D Program of China under Grants No. 2022YFB2804504; Major scientific and technological program of Jilin Province under Grants No. 20210301014GX. Jilin Provincial Development and Reform Commission project (2020C056). Program for Jilin University Science and Technology Innovative Research Team (JLUSTIRT，2021TD-39)

## Author details
1 State Key Laboratory on Integrated Optoelectronics, College of Electronic
Science and Engineering, Jilin University, Changchun 130012, China.
2 Peng Cheng Laboratory, Shenzhen 518000, China.
3 Advance Micro Foundry Pte. Ltd., 11 Science Park Road, Science Park II 117685, Singapore.

## Author contributions
B.C., Y.L. and J.S. led this work. B.C., Y.L., Q.N. and Q.X. performed the experiments and analyzed results. B.C., M.T. and Z.W. built the control system. G.L. support the wafer fabrication process. B.C., Y.L. and J.S. prepared the paper. B.C., Q.X., Y.L., Q.N., G. L. and J.S. revised the paper.

## Conflict of interest
The authors declare no competing interests.

## References
1   Ronchini Ximenes, A. *et al.* A Modular, Direct Time-of-Flight Depth Sensor in 45/65-nm 3-D-Stacked CMOS Technology. *IEEE Journal of Solid-State Circuits* **54**, 3203-3214, doi:10.1109/jssc.2019.2938412 (2019).
2   Morimoto, K. *et al.* Megapixel time-gated SPAD image sensor for 2D and 3D imaging applications. *Optica* **7**, doi:10.1364/optica.386574 (2020).
3   Hutchings, S. W. *et al.* A Reconfigurable 3-D-Stacked SPAD Imager With In-Pixel Histogramming for Flash LIDAR or High-Speed Time-of-Flight Imaging. *IEEE*

*Journal of Solid-State Circuits* **54**, 2947-2956, doi:10.1109/jssc.2019.2939083 (2019).

4   Aflatouni, F., Abiri, B., Rekhi, A. & Hajimiri, A. Nanophotonic coherent imager. *Opt Express* **23**, 5117-5125, doi:10.1364/OE.23.005117 (2015).

5   Fatemi, R., Abiri, B., Khachaturian, A. & Hajimiri, A. High sensitivity active flat optics optical phased array receiver with a two-dimensional aperture. *Opt Express* **26**, 29983-29999, doi:10.1364/OE.26.029983 (2018).

6   Rogers, C. *et al.* A universal 3D imaging sensor on a silicon photonics platform. *Nature* **590**, 256-261, doi:10.1038/s41586-021-03259-y (2021).

7   Li, C., Cao, X., Wu, K., Li, X. & Chen, J. Lens-based integrated 2D beam-steering device with defocusing approach and broadband pulse operation for Lidar application. *Opt Express* **27**, 32970-32983, doi:10.1364/OE.27.032970 (2019).

8   Li, C. *et al.* Monolithic transceiver for lens-assisted beam-steering Lidar. *Opt Lett* **46**, 5587-5590, doi:10.1364/OL.438740 (2021).

9   Zhang, X., Kwon, K., Henriksson, J., Luo, J. & Wu, M. C. A large-scale microelectromechanical-systems-based silicon photonics LiDAR. *Nature* **603**, 253-258, doi:10.1038/s41586-022-04415-8 (2022).

10  Poulton, C. V. *et al.* Long-Range LiDAR and Free-Space Data Communication With High-Performance Optical Phased Arrays. *IEEE Journal of Selected Topics in Quantum Electronics* **25**, 1-8, doi:10.1109/jstqe.2019.2908555 (2019).

11  Poulton, C. V. *et al.* Coherent LiDAR With an 8,192-Element Optical Phased Array and Driving Laser. *IEEE Journal of Selected Topics in Quantum Electronics* **28**, 1-8, doi:10.1109/jstqe.2022.3187707 (2022).

12  Chung, S., Abediasl, H. & Hashemi, H. A Monolithically Integrated Large-Scale Optical Phased Array in Silicon-on-Insulator CMOS. *IEEE Journal of Solid-State Circuits* **53**, 275-296, doi:10.1109/jssc.2017.2757009 (2018).

13  Miller, S. A., Phare, C. T., Chang, Y. C., Ji, X. & Lipson, M. in *CLEO: Applications and Technology.*

14  Wang, P. *et al.* Design and fabrication of a SiN-Si dual-layer optical phased array chip. *Photonics Research* **8**, doi:10.1364/prj.387376 (2020).

15  Wang, P. F. *et al.* Improving the performance of optical antenna for optical phased arrays through high-contrast grating structure on SOI substrate. *Opt Express* **27**, 2703-2712, doi:10.1364/OE.27.002703 (2019).

16  Li, Y. *et al.* Wide-steering-angle high-resolution optical phased array. *Photonics Research* **9**, doi:10.1364/prj.437846 (2021).

17  Xu, W. *et al.* Fully Integrated Solid-State LiDAR Transmitter on a Multi-Layer Silicon-Nitride-on-Silicon Photonic Platform. *Journal of Lightwave Technology* **41**, 832-840, doi:10.1109/jlt.2022.3204096 (2023).

18  Fatemi, R., Khachaturian, A. & Hajimiri, A. A Nonuniform Sparse 2-D Large-FOV Optical Phased Array With a Low-Power PWM Drive. *IEEE Journal of Solid-State Circuits* **54**, 1200-1215, doi:10.1109/jssc.2019.2896767 (2019).

19  Liu, Y. & Hu, H. Silicon optical phased array with a 180-degree field of view for 2D optical beam steering. *Optica* **9**, doi:10.1364/optica.458642 (2022).

20  Hutchison, D. N. *et al.* High-resolution aliasing-free optical beam steering. *Optica* **3**, doi:10.1364/optica.3.000887 (2016).

21  Phare, C. T., Shin, M. C., Sharma, J., Ahasan, S. & Lipson, M. in *CLEO: Science and Innovations.*

22  Na, Q. *et al.* Optical frequency shifted FMCW Lidar system for unambiguous measurement of distance and velocity. *Optics and Lasers in Engineering* **164**, doi:10.1016/j.optlaseng.2023.107523 (2023).

23  Dostart, N. *et al.* Vernier optical phased array lidar transceivers. *Opt Express* **30**, 24589-24601, doi:10.1364/OE.451578 (2022).

24  Chen, B. *et al.* Unidirectional large-scale waveguide grating with uniform radiation for optical phased array. *Opt Express* **29**, 20995-21010, doi:10.1364/OE.427999 (2021).

25  Fang, Q. *et al.* Ultralow Power Silicon Photonics Thermo-Optic Switch With Suspended Phase Arms. *IEEE Photonics Technology Letters* **23**, 525-527, doi:10.1109/lpt.2011.2114336 (2011).

26  Komljenovic, T., Helkey, R., Coldren, L. & Bowers, J. E. Sparse aperiodic arrays for optical beam forming and LIDAR. *Opt Express* **25**, 2511-2528, doi:10.1364/OE.25.002511 (2017).

27  Tao, M. *et al.* A Large-Range Steering Optical Phased Array Chip and High-Speed Controlling System. *IEEE Transactions on Instrumentation and Measurement* **71**, 1-12, doi:10.1109/tim.2021.3139692 (2022).

28  Riemensberger, J. *et al.* Massively parallel coherent laser ranging using a soliton microcomb. *Nature* **581**, 164-170, doi:10.1038/s41586-020-2239-3 (2020).

29  Tang, L., Li, L., Li, J. & Chen, M. Hybrid integrated ultralow-linewidth and fast-chirped laser for FMCW LiDAR. *Opt Express* **30**, 30420-30429, doi:10.1364/OE.465858 (2022).

30  Tang, L. *et al.* Hybrid integrated low-noise linear chirp frequency-modulated continuous-wave laser source based on self-injection to an external cavity. *Photonics Research* **9**, doi:10.1364/prj.428837 (2021).

31  Wang, Z. *et al.* Metasurface empowered lithium niobate optical phased array with an enlarged field of view. *Photonics Research* **10**, doi:10.1364/prj.463118 (2022).

32  Nickerson, M. *et al.* Gallium arsenide optical phased array photonic integrated circuit. *Optics Express* **31**, doi:10.1364/oe.492556 (2023).

33  Wang, J. Y. Heterodyne laser radar-SNR from a diffuse target containing multiple glints. *Applied Optics* **21**, 464-476, doi:10.1364/AO.21.000464 (1982).

# Supplementary Information for

# SiN-on-SOI Optical Phased Array LiDAR for Ultra-Wide Field of View and 4D Sensing

Baisong Chen[1], Yingzhi Li[1], Qijie Xie[2], Quanxin Na[2], Min Tao[1], Heming Hu[1], Ziming Wang[1], Zihao Zhi[1], Xuetong Li[1], Huan Qu[1], Xiaolong Hu[1], Guoqiang Lo[3] and Junfeng Song[1,2,*]

*Corresponding author. Email: songjf@jlu.edu.cn;

**Affiliations:**

[1]State Key Laboratory on Integrated Optoelectronics, College of Electronic Science and Engineering, Jilin University, 130012 Changchun, China.

[2]Peng Cheng Laboratory, 518000 Shenzhen, China.

[3]Advance Micro Foundry Pte. Ltd., 11 Science Park Road, Science Park II, 117685, Singapore.

**This file includes:**

Supplementary Section 1 to Section 8

Table S1

Figures S1 to S9

Reference 1 to 11

## Section 1. Comparison of OPAs in recent years.

Table S1. Comparison of OPA-based beam scanner and OPA-based LiDAR

| Year | Mechanism for wide FoV | Steering FoV | imaging FoV | Reference |
|---|---|---|---|---|
| 2018 | Index-mismatch waveguides | 120° | -[a] | 1 |
| 2019 | 1.65 μm pitch | 56° | 56° | 2 |
| 2020 | Slab-grating | 96° | -[a] | 3 |
| 2021 | Non-uniform antenna array | 140° | -[a] | 4 |
| 2022 | Index-mismatch waveguides and slab-grating | 140° | -[a] | 5 |
| 2022 | 1 μm pitch | 100° | 100° | 6 |
| 2023 | Non-uniform antenna array | 140° | -[a] | 7 |
| **This work** | **Vernier OPA and Wide far-field profile** | **~160°** | **~160°** | |

[a] 3D imaging is not achieved.

## Section 2. Performance of SiN dual-level grating antenna with large aperture

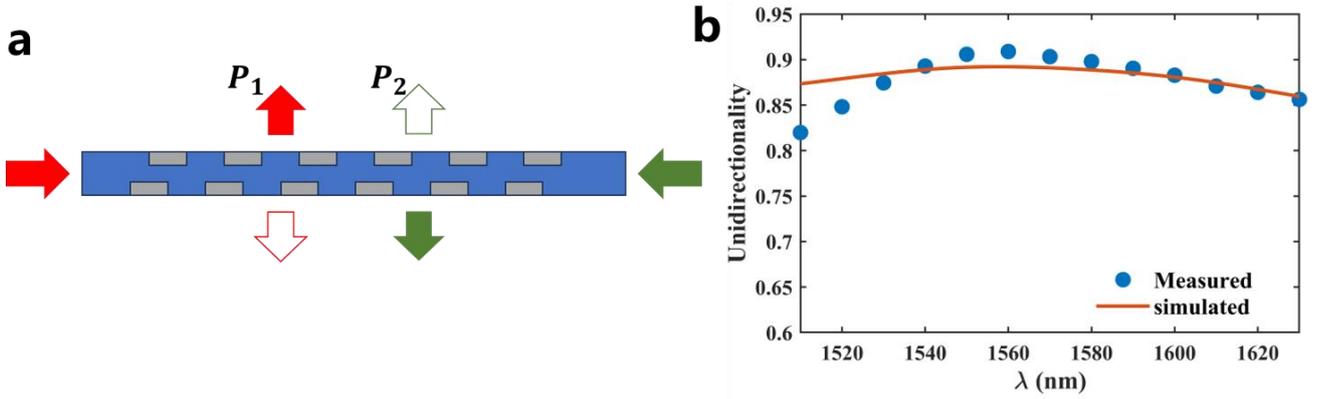

**Fig. S1 Demonstration of silicon nitride dual-level grating antenna with unidirectional emission. a** Schematic for grating antenna directionality test. **b** Measured and simulated results of grating antenna directionality.

Common grating antennas have almost equal emission in the upward and downward direction. However, generally only the power of upward emission can be utilized, resulting in the efficiency of antennas less than 50%. We design a silicon nitride dual-level grating antenna, which achieves high unidirectionality by offsetting the upper and lower grating structures. The testing structure is illustrated in Fig. S1. The emission pattern of antenna varies according to the direction of input light[2,6,8]. When a light is inputted from the left-hand side, the antenna mainly emits the light upward ($P_1$) and partially downward as shown by the red arrow in Fig. S1a. In contrast, when the light is sent from the right-hand side, the main emission direction reverses. In this case, the power of upward emission is $P_2$. We measure the $P_1$ and $P_2$, through coupling light from left and right sides, respectively. Subsequently, the unidirectionality can be calculated by the formula $P_1/(P_1 + P_2)$.

As shown in the Fig. S1b, the measured and simulated unidirectionality varies with the input wavelength within the bandwidth from 1510 nm to 1630 nm. According to the measurement results, the maximum unidirectionality is about 91% in 1560 nm, and the minimum unidirectionality is 82%.

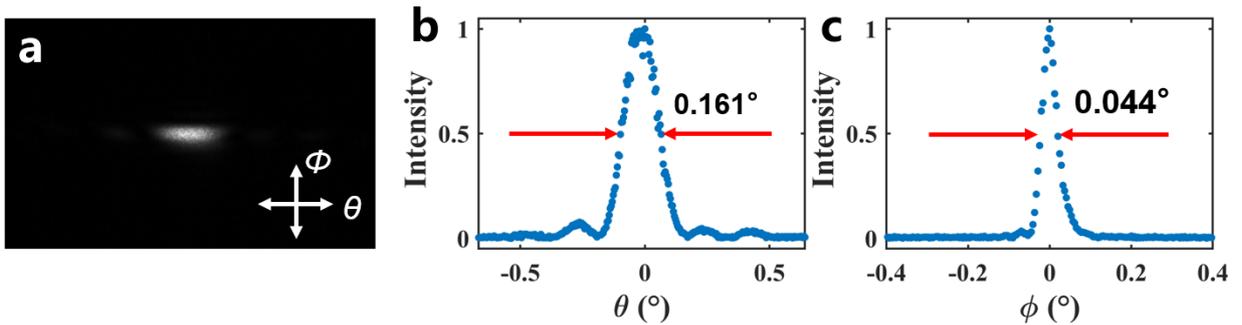

**Fig. S2 Cross section of main lobe at 0°. a** The beam cross-section is measured at 5.8 m away from the fabricated OPA. **b** The beam cross-section in $\theta$ direction. **c** The beam cross-section in $\Phi$ direction

The cross section of main lobe at 0° is measured by infrared camera, as shown in Fig. S2a. The main beam width of fabricated OPA is determined by antenna aperture. The uniform-pitch 128-element grating antenna array with a pitch of 3.8 μm provides a beam with horizontal divergence of 0.161° at the 0° steering angle, as shown in Fig. S2b. The SiN dual-level grating antennas with weak perturbation enable a large aperture, which can provide the beam with vertical divergence of 0.044°, as shown in Fig. S2c.

# Section 3. Simulation of the chained grating antenna with broad far-field profile

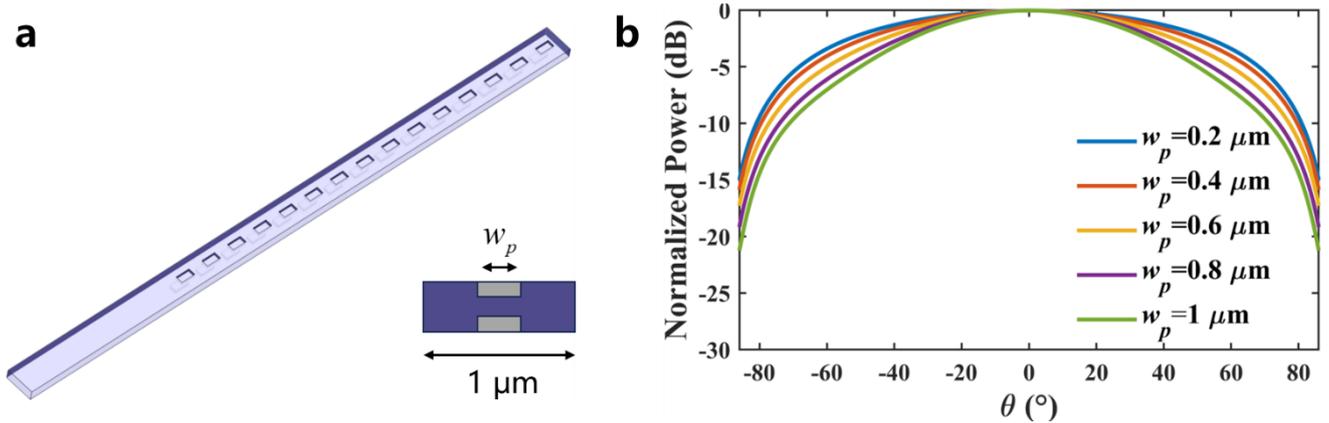

**Fig. S3 Demonstration of dual-level chained grating antenna for broad far field profile. a** Schematic structure of dual-level chained grating antenna, and the inset is the cross-section of the grating. The width of silicon nitride waveguide is 1 μm, and the etching widths on both the upper and lower surface are denoted by $w_p$. **b** The plot of power profile at various etching width $w_p$.

The width of far-field profile is determined by the size of grating antenna[4,9,10]. Silicon with high reflective index difference to cladding has good confinement of guided light, which allows small footprint of antenna to enlarge far-field profile. Hutchison et al. employed a 400 nm thick silicon waveguide for good confinement to broaden far-field profile, and finally demonstrate 80° steering angle. However, thick silicon waveguides impose high requirements on process, and cannot be used to broaden the far-field profile further. We present a SiN chained grating antenna, which broadens the far-field profile effectively by simply reducing the etching size[4]. Schematic structure of dual-level chained grating antenna is shown in Fig. S3a. The width of SiN waveguide is 1 μm, and the etching widths on both the upper and lower surface are denoted by $w_p$. Fig. S3b shows the 3D finite difference time domain (FDTD) simulation result of the far-field profile at different etching apertures $w_p$. It is found that the 3-dB width of far-field profile is increased from 77° to 115° as the $w_p$ decreases from 1 μm to 0.2 μm.

## Section 4. Measurement of beam cross section within a FoV of ~160°

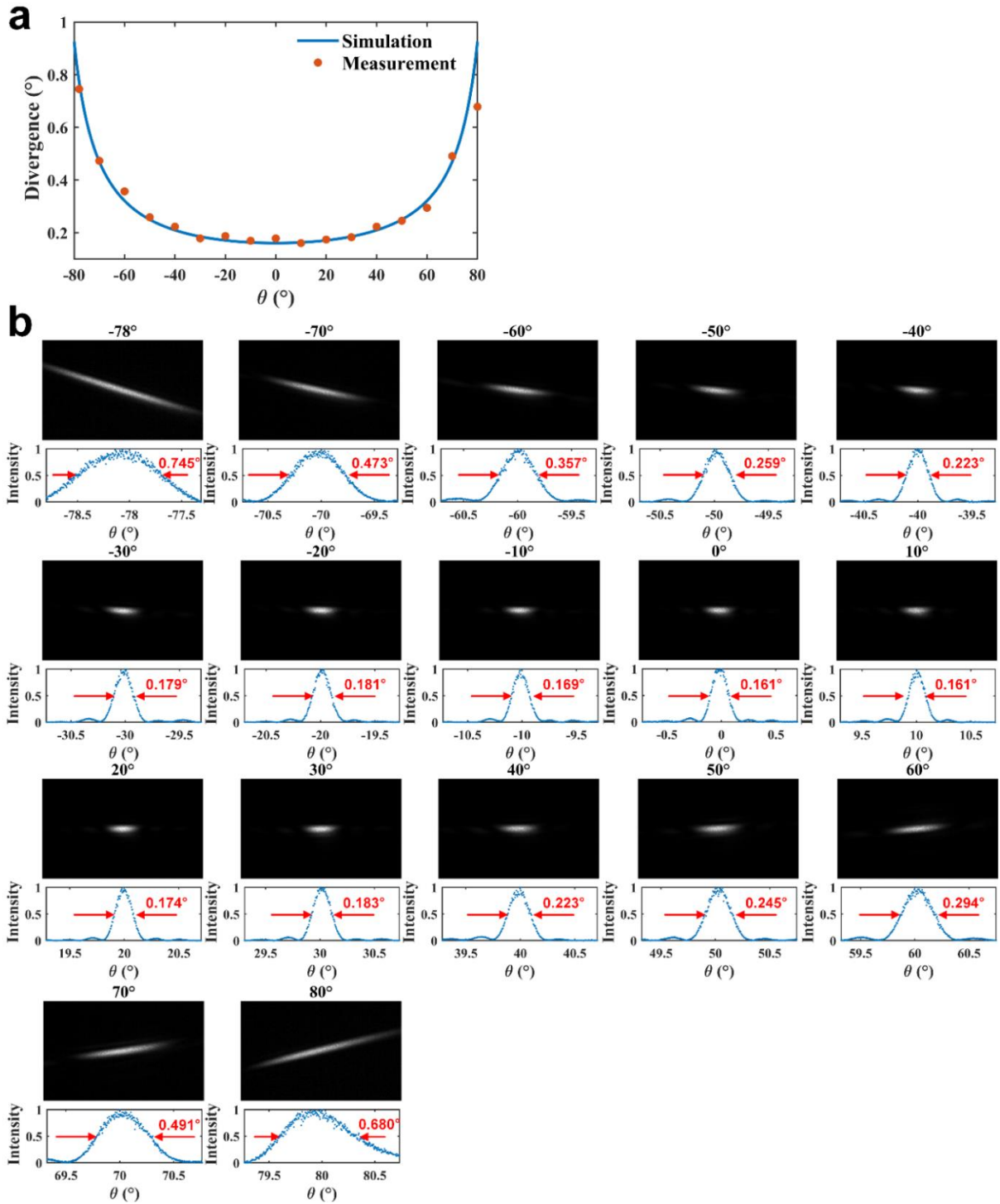

**Fig. S4 Characterization of beam quality in a FoV of ~160°. a** The horizontal divergence of beam in a FoV of ~160°. The blue line represents the simulated result, and the orange dots represent the experimental measurement. **b** Measured beam cross-section within the FoV at 5.8 m distance away from the OPA.

    The main beam width is determined by antenna aperture. The beam width is the narrowest (0.161°) at the 0° and gradually expanded to around 0.7° at the edges of the FoV. The reason for beam width expansion is following. When the beam is emitted at a grazing angle, the effective aperture decreases, resulting in a broadened beam width. The effective way for narrowing the beam width is to scale up the number of OPA elements.

# Section 5. Operation mechanism of Vernier OPA

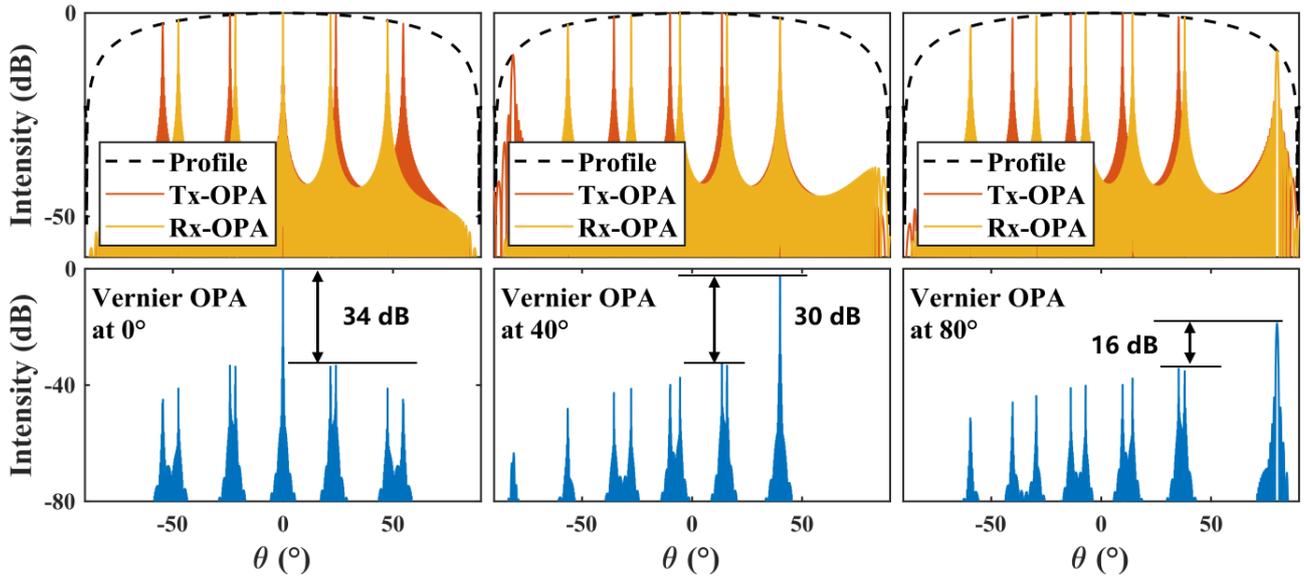

**Fig. S5** Simulated far fields of a single Tx-OPA, a single Rx-OPA (upper branch) and the combined far fields of Vernier OPA (lower branch) at 0°, 40° and 80°, respectively. The black dash curve indicates the wide far-field profile provided by the chained grating antenna.

Considering the wide far-field profile of chained grating antenna, the simulated far fields of Tx- and Rx-OPA are shown in the upper of Fig. S5, and the corresponding far field for Vernier OPA is shown in the lower of the Fig. S5. The resultant far field of Vernier OPA is synthesized by overlapping the far fields of Tx- and Rx-OPA. When the steering angle is 0° and 40°, the sidelobe suppression ratios (SMSR) of Vernier OPA are about 34 dB and 30 dB, respectively. The SMSR still can be maintained over 16 dB even at the extreme angle of 80°.

# Section 6. Experimental verification of Vernier OPA

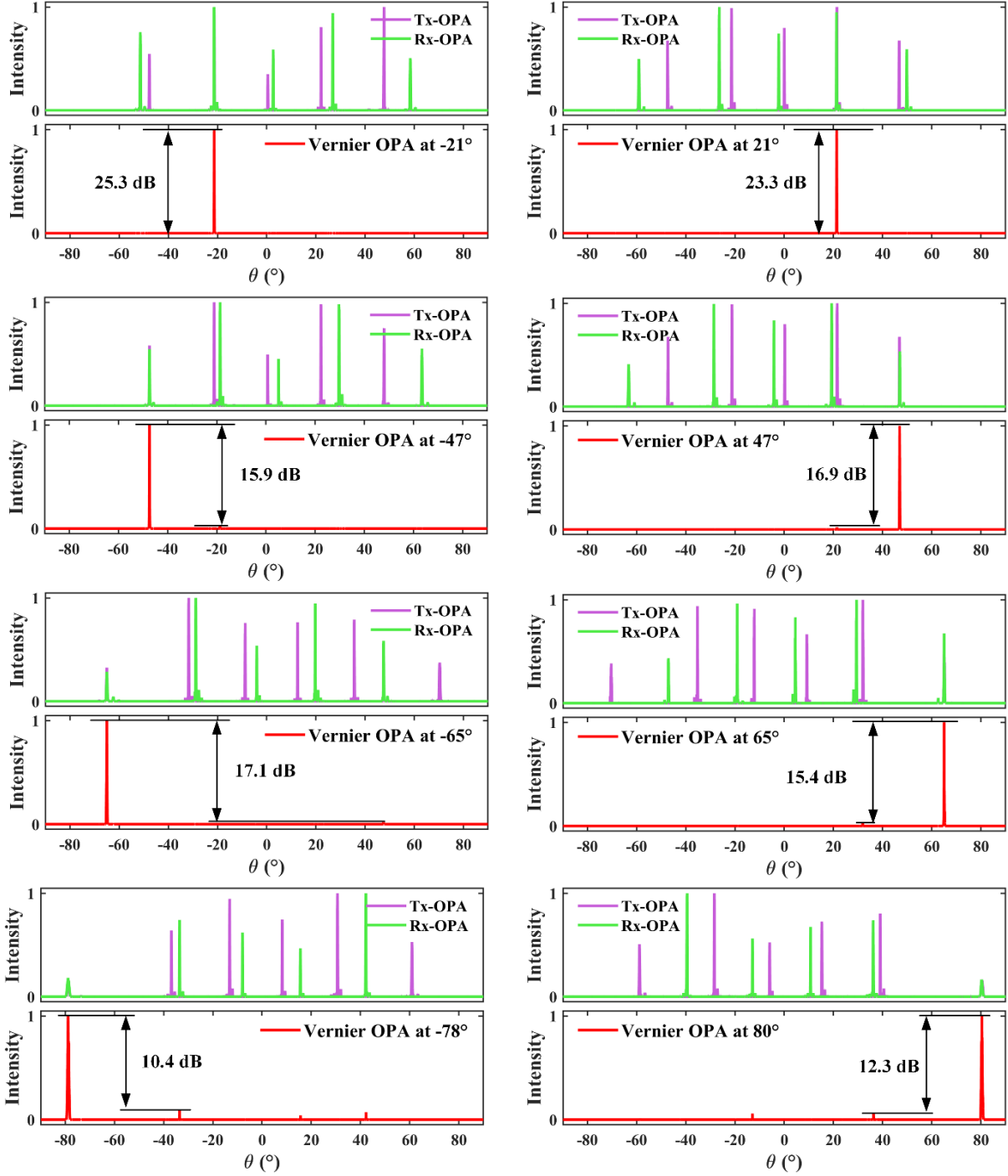

**Fig. S6 Measured results for far fields of Tx-OPA, Rx-OPA (upper) and the corresponding Vernier OPA (lower) within a wide FoV from -78° to 80°.**

By stitching images captured by infrared camera, the far fields of the whole FoV can be measured. The far field of Vernier OPA can be synthesized by overlapping the far fields of Tx-and Rx-OPA. Within a wide FoV from -78° to 80, the far fields of Tx- and Rx-OPA, and corresponding far fields of Vernier OPA are demonstrated in Fig. S6. The side lobe suppression ratio (SMSR) is greater than 10 dB in the full FoV.

## Section 7. The quality of beat signals for ranging within a ~160° FoV

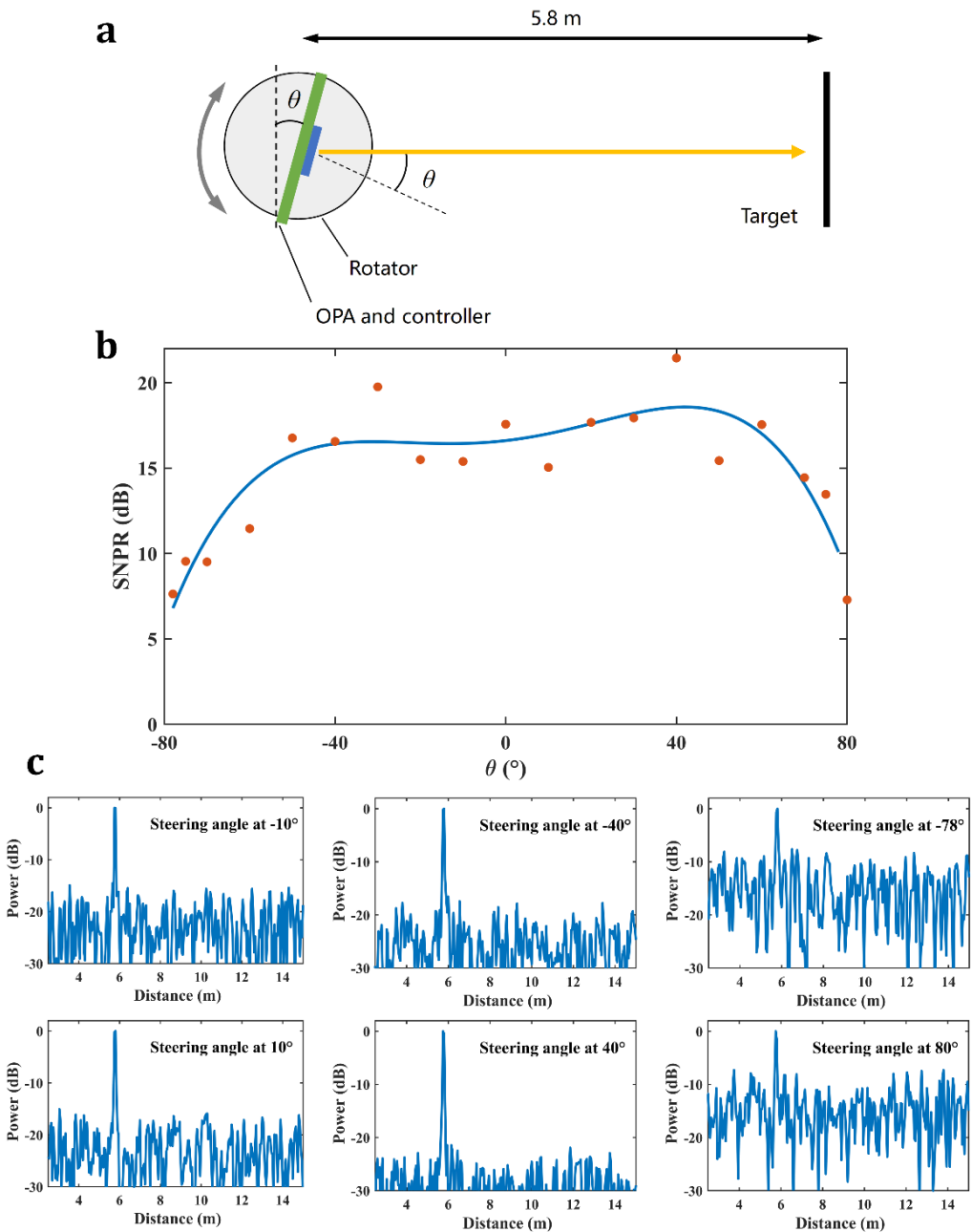

**Fig. S7 The experimental setup and results of beat signals for ranging within the ~160° FoV. a** Experimental setup for ranging at different steering angles given the same distance. **b** The plot for signal-to-noise pedestal ratios (SNPR) of the beat signals against as the steering angle within 160° FoV. The orange dots represent the experimental measurement and the blue line represents the fitting curve. **c** Representative beat signals in the FoV.

As shown in Fig. S7a, the OPA and controller are placed on a rotator, and a target is placed 5.8 m away from the OPA. By rotating the OPA, the quality of ranging signals of different angles within the entire FoV can be fairly compared. The SNPRs of beat signals within a ~160° FoV are measured and demonstrated in Fig. S7b. It is found that due to the wide far-field profile of chained grating antenna, the quality of ranging signals is almost consistent within -60° to 70°. Even at edge of the FoV (-78° and 80°), the SNPRs are still more than 7.5dB. Therefore, the available FoV of proposed OPA-based LiDAR can reach ~160°. Fig. S7c shows the beat signals in the case of the Vernier OPA steered at -78°, -40°, -10°, 10°, 40°, 80°.

## Section 8. Analysis of ranging accuracy of the proposed OPA-based LiDAR

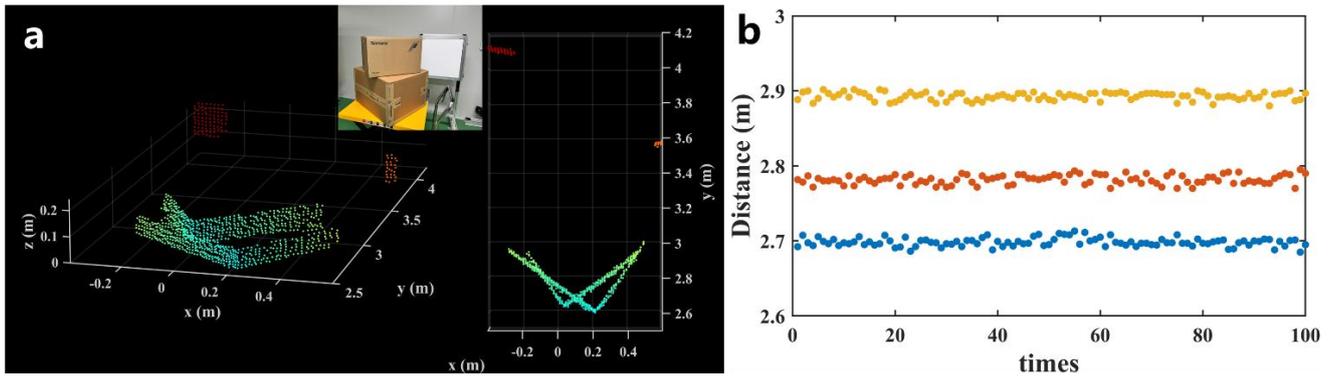

**Fig. S8 Ranging accuracy of the proposed OPA-based LiDAR. a** Point clouds and camera image of stacked cardboard boxes. **b** Ranging results for three positions on the cardboard box via 100-times measurement.

The 3D point clouds of two stacked cardboard boxes and the corresponding real scene are shown in Fig. S8a. The outline of the cardboard boxes can be distinguished clearly. In the projection on the XY plane, it is found that the continuous outlines and two corners of the cardboard boxes are clearly displayed in the point clouds. To verify the ranging accuracy, we measured three positions on the boxes by 100 times respectively. The measured results are shown in Fig. S8b. The standard deviation of the ranging result is 5.1 mm, 5.4 mm and 6.1 mm, respectively. Hence, the ranging accuracy is the average of three ranging standard deviations, which is 5.5 mm. Further linearization of chirp light can improve ranging accuracy[11].

## Section 9. A record wide scene 3D imaging enabled by the proposed OPA-based LiDAR

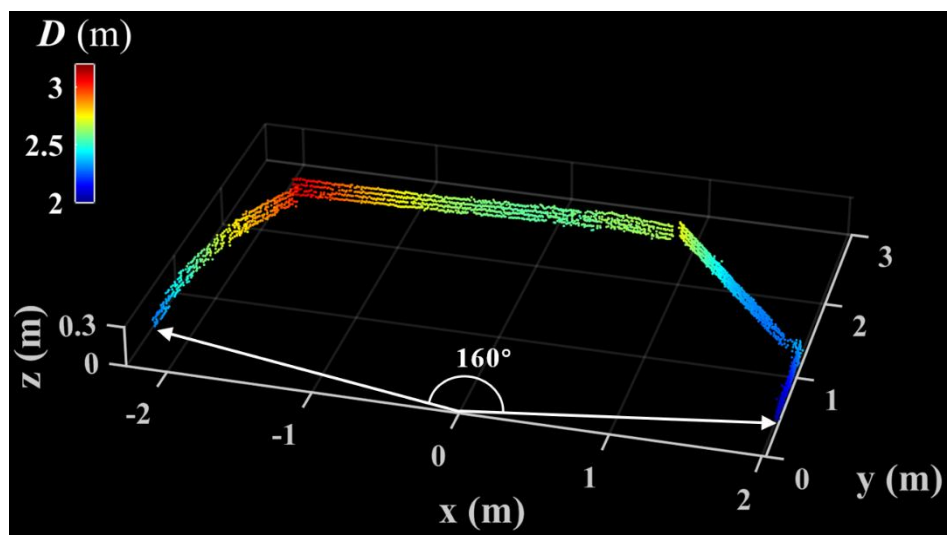

**Fig. S9 3D point-clouds image with a 160° FoV of by using the proposed OPA-based LiDAR.**

We present an OPA-based LiDAR to achieve a 3D imaging with a remarkable FoV. Fig. S9 shows the experimentally measured 3D point clouds with a record wide FoV of 160°. For the proof of concept, the image consists of 4 rows and each row concludes 620 horizontal resolvable points. The distance between the target scene and the OPA chip is around 2 meters, which is well above the 1.5-m blind zone. The blind zone of the OPA-based LiDAR system is mainly attributed to the low-frequency noise generated from the RF generator module. The noise performance can be improved by optimizing the RF module. Besides, the 1.5-m blind zone poses stringent requirement on the space to achieve wide-scene imaging. Due to the limited space of the laboratory, a preliminary result for high-resolution 3D image is restricted to 135° horizontal FoV as presented in the manuscript. Afterward, we have further demonstrated 3D imaging with 160° horizontal FoV on a temporary room for the proof-of-concept experiment.

## Reference

1. Phare, C. T., Shin, M. C., Sharma, J., Ahasan, S. & Lipson, M. in *CLEO: Science and Innovations.*
2. Poulton, C. V. *et al.* Long-Range LiDAR and Free-Space Data Communication With High-Performance Optical Phased Arrays. *IEEE Journal of Selected Topics in Quantum Electronics* **25**, 1-8, doi:10.1109/jstqe.2019.2908555 (2019).
3. Wang, P. *et al.* Design and fabrication of a SiN-Si dual-layer optical phased array chip. *Photonics Research* **8**, doi:10.1364/prj.387376 (2020).
4. Li, Y. *et al.* Wide-steering-angle high-resolution optical phased array. *Photonics Research* **9**, doi:10.1364/prj.437846 (2021).
5. Liu, Y. & Hu, H. Silicon optical phased array with a 180-degree field of view for 2D optical beam steering. *Optica* **9**, doi:10.1364/optica.458642 (2022).
6. Poulton, C. V. *et al.* Coherent LiDAR With an 8,192-Element Optical Phased Array and Driving Laser. *IEEE Journal of Selected Topics in Quantum Electronics* **28**, 1-8, doi:10.1109/jstqe.2022.3187707 (2022).
7. Xu, W. *et al.* Fully Integrated Solid-State LiDAR Transmitter on a Multi-Layer Silicon-Nitride-on-Silicon Photonic Platform. *Journal of Lightwave Technology* **41**, 832-840, doi:10.1109/jlt.2022.3204096 (2023).
8. Wang, Z. *et al.* Wide field of view optical phased array with a high-directionality antenna. *Opt Express* **31**, 21192-21199, doi:10.1364/OE.492317 (2023).
9. Hutchison, D. N. *et al.* High-resolution aliasing-free optical beam steering. *Optica* **3**, doi:10.1364/optica.3.000887 (2016).
10. Sun, J., Timurdogan, E., Yaacobi, A., Hosseini, E. S. & Watts, M. R. Large-scale nanophotonic phased array. *Nature* **493**, 195-199, doi:10.1038/nature11727 (2013).
11. Tang, L., Li, L., Li, J. & Chen, M. Hybrid integrated ultralow-linewidth and fast-chirped laser for FMCW LiDAR. *Opt Express* **30**, 30420-30429, doi:10.1364/OE.465858 (2022).